\documentclass[vecphys]{svmult}

\usepackage{graphicx}        
\usepackage[bottom]{footmisc}

\begin{document}

\title*{Eta Carinae and Nebulae Around Massive Stars: Similarities to
  Planetary Nebulae?}
\titlerunning{Massive Star Nebulae}

\author{Nathan Smith}
\authorrunning{N. Smith}
\institute{University of California, Berkeley, 601 Campbell Hall,
  Berkeley, CA 94720, USA; \texttt{nathans@astro.berkeley.edu}}
\maketitle

\begin{abstract}

I discuss some observational properties of aspherical nebulae around
massive stars, and conclusions inferred for how they may have formed.
Whether or not these ideas are applicable to the shaping of planetary
nebulae is uncertain, but the observed similarities between some PNe
and bipolar nebulae around massive stars is compelling.  In the
well-observed case of Eta Carinae, several lines of observational
evidence point to a scenario where the shape of its bipolar nebula
resulted from an intrinsically bipolar explosive ejection event rather
than an interacting winds scenario occurring after ejection from teh
star.  A similar conclusion has been inferred for some planetary
nebulae.  I also briefly mention bipolar nebulae around some other
massive stars, such as the progenitor of SN~1987A and related blue
supergiants.

\end{abstract}

\section{Introduction: Massive Stars and PNe}

Although this is a meeting on aspherical planetary nebulae (PNe),
which are the descendants of intermediate/low-mass stars, I'd briefly
like to shift gears and discuss massive stars.  In the nebulae around
massive stars, like PNe, we see a wide variety of non-spherical
geometries with a common theme of bipolar shapes in the ejecta. For
most of the more stunning examples of massive star nebulae, one can
usually find a PN with nearly identical appearance, at least
superficially.  Some of the more familiar comparisons are $\eta$~Car
to Hb~5, Mz~3, or even the Red Rectangle (Soker 2007), as well as the
similar multiple rings seen around SN~1987A, the luminous blue
variable (LBV) star HD~168625 (Smith 2007), the Red Square (Tuthill \&
Lloyd 2007), and He~2-104 (Corradi et al. 2001), or the peculiar
double rings around RY Scuti (Smith et al.\ 2002) and Abel 14, to name
a few.  Not surprisingly, discussions of the shaping mechanisms for
massive stars and PNe share common themes: binaries/mergers vs.\
interacting winds vs.\ rotating ejections.  For massive stars, though,
magnetic fields still seem to be mostly taboo for the time-being.

Also, like lower mass stars during the AGB phase, it seems to be the
case that massive stars shed most of their mass in a brief post-MS
evolutionary phase, either as a RSG or an LBV.  This was not always
thought to be the case: for very massive stars, the relative
importance of LBV eruptions vs.\ steady winds has been appreciated
fairly recently because of the revised lower mass-loss rates estimated
for O stars on the main sequence, and because of the very high masses
of LBV nebulae (see Smith \& Owocki 2006).

Despite vastly different amounts of mass and energy, the compelling
similarities between massive star nebulae and PNe make it worthwhile
to ask if conclusions gleaned from massive stars can inform the
shaping mechanisms of PNe, and vice versa.  In the interest of being
provocative, then, I'll mention some results for Eta Carinae and a few
other massive stars that have been studied in detail, which challenge
some familiar ideas developed from the study of PNe.  But first, some
general comments on winds.

\section{Interacting Winds Scenarios?}

A fast wind sweeping into a slower and denser wind is a natural avenue
to pursue for shaping nebulae, and such models have had varying
degrees of success in reproducing PNe shapes (there are dozens of
potential references to cite here, including many in these
proceedings).  A similar process may occur in {\it some} massive stars
if they pass through a very slow-wind phase as a RSG and then evolve
through a faster wind phase as a BSG/LBV or a Wolf-Rayet (WR) star.
This can and does produce a wind-blown bubble around the WR star in
some cases.

A key point, though, is that in the case of massive stars we have some
problems if we want interacting winds to account for most bipolar
nebulae. First, the slower nebulae around RSGs and yellow hypergiants
are generally not axisymmetric.  They are often asymmetric or chaotic,
but they almost never have clear signs of organized axisymetry (see,
e.g., VY CMa [Smith et al.\ 2001]; NML Cyg [Schuster et al.\ 2006];
IRC+10420 [Humphreys et al.\ 2002]).  Second, the resulting wind-blown
bubbles around WR stars are {\it NOT} bipolar!\footnote{Some WR
nebulae are a little egg-shaped, but we don't see any with pinched
waists.  Also, I'm not including the ``pinwheels'' and related
phenomena (see, e.g., Tuthill et al.\ 1999) around dust-producing WC
stars, which represent a very different type of ``interacting winds''.}
This lack of axisymmetry occurs despite the apparent fact that massive
stars have high binary fractions.  What does that mean?  If binary
mergers and jets blown by binaries are dominant shaping mechanisms,
shouldn't we see signs of axisymmetry at all stages?  Why is it the
case that the only bipolar/pinched-waist nebulae around massive stars
are those seen around blue supergiants such as LBVs and B[e]
supergiants?

A critical point, I think, is that {\it in these blue supergiants,
their escape speeds, observed ejecta/wind speeds, and surface rotation
speeds are all comparable}.  They are around 100--200 km s$^{-1}$, as
opposed to 10--20 km s$^{-1}$ for RSGs and 1000-2000 km s$^{-1}$ for
WR stars.  I suspect that this is an important clue that for many
nebulae around massive stars, intrinsically aspherical ejection from
the surface of a rotating star is a prime agent in shaping their
nebulae (see Smith \& Townsend 2007).  In the context of interacting
winds, then, I suspect that a very interesting avenue to pursue is an
aspherical fast wind interacting with a slow spherical wind or thin
shell.  In fact, due to the sporadic nature of episodic mass loss from
massive stars, thin shells rather than steady winds is probably where
most of the circumstellar mass resides.

Now, that discussion of interacting winds was for massive stars that
go through a slow-wind RSG phase followed by a fast wind phase in
their evolution...but that only occurs up to initial masses of about
40 M$_{\odot}$.  Stars with higher initial mass (like Eta Car and most
LBVs) never pass through a RSG phase.  So for these stars, {\it
ejection as an LBV is the slowest the wind speeds ever get...but it is
precicesly those LBV nebulae that are observed to be bipolar}.  How
can this be?  This means that they can't be shaped by interacting
winds, because a slow dense wind blowing into a faster rarefied wind
doesn't produce much interaction.  Instead, the slow dense wind that
follows the fast wind needs to be shaped on its own; this is discussed
and amplified below.  Despite the lack of interaction between the fast
and slow wind, they produce shapes very similar to some PNe --- a fact
worth considering.

\section{Eta Carinae}

Eta Carinae is a key object for trying to understand the shaping of
bipolar nebulae, partly because is it bright and so well-observed, and
partly because we have caught it so soon (only 160 yr) after its
violent mass ejection, before its shape has been corrupted by
interaction with the ISM.  Despite its status as the most luminous and
most massive star known, there is considerable overlap with some
topics in PNe research, as I will highlight here.

\subsection{Energy and Momentum}

Studies of the mass, kinematics, and detailed structure have led to
the following basic results (summarized from Smith 2006; Smith et al.\
2003):

1.  The nebula follows a Hubble-like expansion law, with the same age
    for the equatorial and polar ejecta.

2.  The walls of the nebula are very thin, indicating that the
    duration of mass ejection was less than 10\% of the time elapsed
    since ejection.

3.  Essentially all the mass is in the thin molecular shell, formed
    from material ejected by the star in the outburst, not in swept-up
    material.

4.  The large mass, momentum, and kinetic energy came from a single
    exlposive event, and could not have been driven by radiation
    pressure alone or by the stellar wind that has blown after the
    eruption.

All these clues point to a single violent bipolar explosion that
ejected the nebula seen today.  Interestingly, all these same basic
conclusions were inferred by Alcolea et al.\ (2007) from a similar
detailed study of the PN M~1-92.

In the case of Eta Carinae, though, the difficulties for an
interacting winds scenario are compounded further.  The mass as a
function of latitude has been measured in the bipolar lobes around Eta
Car, showing that most of the mass comes from high latitudes near the
pole (Smith 2006).  This rules-out the familiar type of
interacting-wind scenario where a spherical wind plows into a disk or
torus (e.g., Frank et al.\ 1995), because in that scenario, the
pinched waist is essentially the result of mass loading at low
latitudes (note that the bipolar nebula formed in a merger model is a
variation of this).\footnote{By the way, note as well that the disk
seen in {\it HST} images of Eta Car is {\it not} the agent responsible
for pinching the waist of the bipolar nebula, because it is the same
age or younger.}  Similarly, a different type of interacting winds
scenario where a fast aspherical wind\footnote{The present-day wind
appear to be bipolar on size scales smaller than the binary separation
(see Smith et al.\ 2003; van Boekel et al.\ 2003).}  plows into a
slower wind doesn't work either (Frank et al.\ 1998; Gonzalez et al.\
2004).  This is because we can observe the stellar wind that has been
blowing after the 19th century eruption, potentially inflating and
shaping the nebula.  However, it is about 1000 times too weak to shape
the polar lobes (like a light breeze blowing on a brick wall), and
besides, the post-outburst wind speed is almost the same as that of
the nebula, so the winds are not interacting anyway!  There seems to
be little way to escape the conclusion that the bipolar shape of the
Homunculus nebula around Eta Car resulted from an intrinsically
bipolar ejection by the star itself, and not from any sort of
interacting winds scenario.  A possible avenue to pursue is discussed
after the next section.

\subsection{Double-Shell Structure}

I'd like to diverge for a moment to talk about the detailed ionization
structure in the walls of the nebula around Eta Car, as opposed to its
overal bipolar shape, where additional similarities to some PNe can be
seen.  Eta Car's nebula has a distinct double-shell structure, with a
thin outer shell composed of molecular gas and cool dust, and a
thicker inner shell of partially-ionized atomic gas and warmer dust
(Smith 2006; Smith et al.\ 2003).  In high-resolution spectra and
images of H$_2$ 2.122 $\mu$m and [Fe~{\sc ii}] 1.644 $\mu$m, this
structure is almost identical to that seen in some PNe, most notably
in M~2-9 (Hora \& Latter 1994; Smith et al.\ 2005).  These near-IR
H$_2$ and [Fe~{\sc ii}] emission lines are usually taken as signposts
for shock excitation (Shull \& Hollenbach 1978) because they are seen
in supernova remnants, and the double-shell structure is reminiscent
of a forward/reverse shock structure that one might expect for
interacting winds (e.g., Chevalier 1982).

However, they can also arise from dense atomic and molecular gas that
is heated radiatively in a dense PDR (e.g., Sternberg \& Dalgarno
1989).  Using CLOUDY simulations, Smith \& Ferland (2007) demonstrated
that the observed IR emission tracers, ionization structure, and the
observed dust temperatures can arise naturally from radiative heating
if the two shells contain roughly the amount of mass inferred from
studies of the dust (Smith et al.\ 2003).  In fact, in the case of Eta
Car, radiative heating dominates the energy budget compared to shock
heating.  This is comforting, because as noted earlier, the
post-eruption wind speed is very similar to that of the nebula ejected
in the eruption, so there is little reason to expect a strong shock
anyway.  So, in Eta Car, it seems clear that the observed ionization
structure arises from radiative excitation, not shocks.  If this is
not true for M~2-9, then the almost identical ionization structure is
quite a coincidence, especially since the spectra of the central
objects are so similar as well (Balick 1989).

\subsection{How to Get Bipolar Lobes and a Disk}

Observations of the bipolar nebula around Eta Car seem to dictate that
it did not arise as a result of an interacting winds scenario, but
instead, from an intrinsically bipolar wind or explosion.  In other
words, gas was launched from the surface of the star imprinted with
the basic bipolar shape seen today.

The present-day, post-eruption wind of Eta Car is also bipolar in
shape with a speed comparable to that of the nebula (Smith et al.\
2003), although it is much weaker than the mass-loss rate during the
19th century eruption that made the nebula.  Its almost as if the
present-day bipolar wind density was simply ``cranked-up'' by a factor
of 1000 during the outburst, maintaining the same basic speed and
shape (e.g., Dwarkadas \& Owocki 2002).  If the bipolar nebula was
created by some other external mechanism (such as jets blown by
accretion onto a companion; see Soker, these proceedings) then it is a
remarkable coincidence that the wind shape and speed so closely match
the present-day properties of the primary star's wind.  On the other
hand, if the primary star ejected that material, then it is not such a
coincidence at all.

In the case of an intrinsically bipolar ejection, what determines the
resulting shape of a nebula is not the {\it density} as a function of
latitude, but the {\it speed} as a function of latitude.  When mass is
driven off the surface of a star, it is a general property that this
material leaves at nearly the star's surface escape speed.  This is
why RSGs with large radii have slow winds, and compact WR stars have
very fast winds.  Now, if a star is rotating fast enough to
significantly modify the effective gravity at the equator (i.e. a
non-negligible fraction of the critical rotation velocity), then the
star's escape speed will vary with latitude, being faster at the poles
and slower at the equator.  Because of this simple effect, the {\it
default} shape we should expect for material driven from the surface
of a rotating star is a bipolar nebula.  Admittedly, for this effect
to shape the wind, the rotation speeds should be comparable to the
wind speed, but this is indeed the case for blue supergiants, as noted
earlier.  The degree to which the waist is pinched depends on how
close the star is to critical rotation.  In some cases, such models
can also make an equatorial disk like that seen in Eta Car.

Smith \& Townsend (2007) described this type of model in detail, with
test particles launched from the surface of a rotating star, following
simple ballistic trajectories thereafter.  They showed that it could
account for the shape and speed of the polar lobes of Eta Car, as well
as the basic properties of its peculiar equatorial disk.  Again, this
is the simplest, default shape one should expect for ejection from a
rotating object.  Smith \& Townsend (2007) noted that it may have
application to some PNe as well.  Please see that paper for further
details.  Matt \& Balick (2004) present a somewhat different intrinsic
shaping model for the present-day stellar wind and disk involving MHD
effects; this type of mechanism might also be relevant if the magnetic
field was strong enough during the outburst.

\section{SN~1987A, Rings, and Mergers}

The triple ring system observed around SN~1987A inspired a great deal
of theoretical work on interacting winds as a potential explanation
for its equatorial ring and bipolar ejecta, as well as the formation
of bipolar nebulae in general.  The basic favored picture is that the
star had a blue loop, with a fast BSG wind pushing into a slower RSG
wind.  The bipolar shape and equatorial ring could arise if that RSG
wind had denser material near the equator (Blondin \& Lundqvist 1993;
Martin \& Arnett 1995), but in order for that to happen, the RSG
needed to have an extra source of angular momentum, such as a binary
merger event (Collins et al.\ 1999).  This has evolved into a complex
model that gives an impressive fit to the observed structure of the
nebula (Morris \& Podsiadlowski 2007) seen in {\it HST} images.

While this view is the result of considerble effort and thought, I
wish to note a few flies in the ointment, which suggest that the
merger model for SN~1987A might not be the final word, and may need to
be revisited.

1.  A merger model followed by a transition from a RSG to BSG requires
that these two events be synchronized with the supernova event itself,
requiring that the best observed supernova in history happens to be a
rare event.

2.  After the RSG swallowed a companion star and then contracted to
become a BSG, it should have been rotating at its critical breakup
velocity.  Even though pre-explosion spectra (Walborn et al.\ 1989) do
not have sufficient resolution to measure line profiles,
Sk--69$^{\circ}$202 showed no evidence of rapid rotation (e.g., like a
B[e] star spectrum).  Instead, Sk--69$^{\circ}$202 had the spectrum of
an entirely normal B3 supergiant.

3.  Particularly troublesome is that this merger and RSG/BSG
transition would need to occur twice. From an analysis of light echoes
for up to 16 yr after the supernova, Sugerman et al.\ (2005)
have identified a much larger bipolar nebula with the {\it same axis
orientation} as the more famous inner triple ring nebula.  If a merger
and RSG/BSG transition are to blame for the bipolarity in the
triple-ring nebula, then what caused it in the older one?

Now, these points may seem silly at first, they are hard to reconcile
with the merger model. The last one, in particular, could even be
considered to be a strong rebuke.  Given that more luminous blue
supergiants can eject intrinsically bipolar nebulae without resorting
to RSG/BSG transitions or mergers, could SN~1987A's nebula be the
result of a massive star ejection instead, like an LBV?

In addition to these problems with the specific case of SN~1987A
itself, we also need to take into account the growing number of
observed nebulae around massive stars with rings similar to SN~1987A.
Is there any evidence that they also formed from mergers?  The example
most people are familiar with is Sher~25, in the massive cluster
NGC~3603 in our Galaxy (see Brandner et al.\ 1997).  It has an
equatorial ring with the same physical radius as that of SN~1987A, and
it also has bipolar lobes.  Yet, studies of the central star in Sher
25 show that its abundances indicate that {\it it has not gone through
a RSG phase} (Smartt et al.\ 2002).  The newly discovered SBW1 in
Carina also has a 0.2 pc radius identical to the equatorial ring
around 87A; its central star has about the same luminosity as that of
the progenitor of SN~1987A, but its ring has Solar N abundances, so it
has also {\it not been through a RSG phase} (Smith et al.\ 2007).
Finally, there's the triple-ring nebula around the LBV star HD~168625
(Smith 2007), which appears to be almost identical to that of
SN~1987A.  While this star could indeed be a post-RSG because it is
mildly N-rich, it is an LBV -- stars well known for their unstable
bipolar mass ejections.  Furthermore, Smith (2007) argued that if the
triple ring nebula of HD~168625 is coeval (polar rings have the same
age as the equator, as in SN1987A), then they could not have been
ejected in a RSG phase because they would be far too fast.

So while the progenitor of SN~1987A may very well have passed through
the RSG phase needed for the merger model, two of its twins did not,
and a third is an LBV, where the shell was likely created in an LBV
ejection.  If a binary merger in the RSG phase really is required to
make the ring around SN~1987A, how and why did at least three other
objects make {\it nearly identical} nebulae in a different way?  At
the very least, this implies that a merger model is not the only
viable option.  Keep in mind that their central stars still exist and
are easy to observe, yet there is currently no evidence that they are
post-merger products.  In fact, they appear as fairly normal blue
supergiants; like the progenitor of SN~1987A, these other rings do not
appear to be unusually rapid rotators as one would expect for a
post-merger product (their inclinations are known from the equatorial
ring nebulae).  If further studies reveal that any of them survive
today as close binaries (binaries that have not yet merged, so that
the observed rings are obviously not from a merger event), it will
critically wound the merger hypothesis for SN~1987A.

{\bf Acknowledgements.} I would like to thank the conference organizers
for the invitiation to speak and for generous partial financial
support that made my trip possible.

\end{document}